# Highly sensitive photodetector based on two-dimensional ferroelectric semiconducting β-InSe/graphene heterostructure

Jialin Li[1], Yuzhong Chen[2], Yujie Li[2], Haiming Zhu[2], Linjun Li[1,3]*

[1]State Key Laboratory of Modern Optical Instrumentation, College of Optical Science and Engineering, Zhejiang University, Hangzhou 310027, China
[2]Center for Chemistry of High-Performance & Novel Materials, Department of Chemistry, Zhejiang University, Hangzhou 310027, China
[3]Intelligent Optics & Photonics Research Center, Jiaxing Research Institute Zhejiang University, Jiaxing 314000, China

*Email: lilinjun@zju.edu.cn

**ABSTRACT**
2D ferroelectric β-InSe/graphene heterostructure was fabricated by mechanical exfoliation, and the carrier dynamics crossing the heterostructure interface has been systematically investigated by Raman, photoluminescence and transient absorption measurements. Due to the efficient interfacial photo excited electron transfer and photogating effect from trapped holes, the heterostructure devices demonstrate superior performance with maximum responsivity of $2.12\times10^4$ A/W, detectivity of $1.73\times10^{14}$ Jones and fast response time (241 µs) under λ = 532 nm laser illumination. Furthermore, the photo responses influenced by ferroelectric polarization field are investigated. Our work confirms ferroelectric β-InSe/graphene heterostructure as an outstanding material platform for sensitive optoelectronic application.

Photodetectors based on two-dimensional (2D) materials and their heterostructures have attracted great attention for the past decades, various devices have been developed to achieve high performance[1-7]. Recently, controlling 2D ferroelectric order has shown a new way to improve the photo response. For example, enhanced bulk photovoltaic effect was reported in 2D ferroelectric $CuInP_2S_6$, with two orders enhancement of the photocurrent associated to its room-temperature polar order[8]. The polarization field induced by ferroelectric order separates photogenerated electron–hole pairs and further efficient transfer. However, it has low photo responsivity (~10 mA/W) when worked as photoconductive mode, due to its large bandgap (2.96 eV) with very low mobility[9]. In addition, to achieve a higher responsivity of 2D semiconductors, previous work chose the ferroelectric poly (vinylidene fluoride-trifluoroethylene) (P(VDF-TrFE)) copolymer films to tune the carrier transport, which needs relatively complex structure and high gating voltage[10, 11]. Another recent work reported the ferroelectric d1T-$MoTe_2$ based vertical structure for photodetector application[12], however it needs additional intensive laser irradiation process and the responsivity (~ 853 A/W) needs to be improved. While for β-InSe, as a direct bandgap 2D material with relatively high mobility up to 1000 $cm^2V^{-1}s^{-1}$ at room temperature[13], has been demonstrated to possess voltage-tunable in-plane and out-of-plane sliding ferroelectricity with metal electrodes[14, 15], also photodetector based on β-InSe with metal electrodes has been shown responsivity up to 194 A /W[16], therefore it is desirable to achieve better performance by developing its heterostructure with other material and controlling its ferroelectric order. However, the optoelectronic application of ferroelectric β-InSe heterostructure and ferroelectricity effect on the photodetector performance are still unexplored.

In this work, graphene/β-InSe/graphene heterostructure (GIGHS) photodetector devices were prepared by mechanical exfoliation (MF), followed by all-dry transfer process to improve sample quality and interfacial contact. Under a λ = 532 nm laser illumination, the devices demonstrate advanced performance with high responsivity of $2.17\times10^2$ A/W, detectivity of $3.78\times10^{13}$ Jones, fast response time (314 µs) (In-plane configuration), and $2.12\times10^4$ A/W, detectivity of $1.73 \times 10^{14}$ Jones, fast response time (241 µs) (Out-of-plane configuration). Furthermore, the influence of ferroelectricity on the photodetector performance was investigated by voltage poling, which is proved to improve the detectivity by six times.

Our β-InSe crystals are obtained commercially (from SixCarbon Technology, China). The graphene and InSe nanosheets are prepared from the bulk crystals on a 285 nm-thick $SiO_2$-coated silicon substrate by mechanical exfoliation (ME), the typical optical microscope (OM) image of exfoliated β-InSe with different thickness is shown in Fig. 1(a). Fig. 1(b) and 1(c) displays that the corresponding Raman and photoluminescence (PL) spectrum of the β-InSe flake with different thickness. As the thickness decreases, the interlayer coupling is suppressed, therefore the intensity of Raman peak decreases with the decrease of the sample thickness. The PL response decreases and undergoes a blue-shift with decreasing material thickness, which is due to the quantum confinement effect caused by the decrease of the thickness of the material, corresponding to the direct bandgap-indirect bandgap transition, similar to the report[17]. Also, the Piezo-response force microscopy (PFM) is used to confirm the ferroelectricity of β-InSe. As shown in Fig. 1(d), the PFM amplitude and phase hysteresis loops were observed (~180 degrees phase switching), indicating the out-of-plane ferroelectricity of the β-InSe nanosheet.

To probe carriers' transport behavior of the β-InSe/graphene heterostructure (IGHS), we fabricated the sample on quartz substrate by dry transfer method[18], the corresponding OM image is shown in Fig. 2(a). Fig. 2(b) depicts the Raman spectrum of the sample. All the Raman peaks of graphene and β-InSe are identified in the overlapped heterostructure region, demonstrating the good coupling between these two materials. Moreover, PL is significantly quenched in the heterostructure area as shown in Fig. 2(c), indicating the efficient electron–hole separation and charge transfer process at the heterostructure interface. Furthermore, to better understand how carriers transport across the IGHS interface, a micro-area pump–probe technique is employed. Under excitation by a 2.06 eV, ~130 μJ cm$^{-2}$ pump pulse, the transient absorption (TA) kinetics of β-InSe and IGHS is shown in Fig. 2(d), which B exciton (512 nm) of β-InSe is selected as the probe wavelength[add reference]. Fitting the kinetics of them by using biexponential function, the relaxation time of IGHS is determined to be $\tau_1 = \tau_2 = 39.3$ ps, which is much faster than β-InSe individual ($\tau_1$=70.4 ps and $\tau_2$ = 1179.2 ps). The significantly suppressed relaxation time of IGHS is due to the fast interfacial charge transfer from β-InSe and Graphene. Therefore, it is benefit for realizing fast photodetection, due to the significantly suppressed recombination of photogenerated charge carriers in β-InSe.

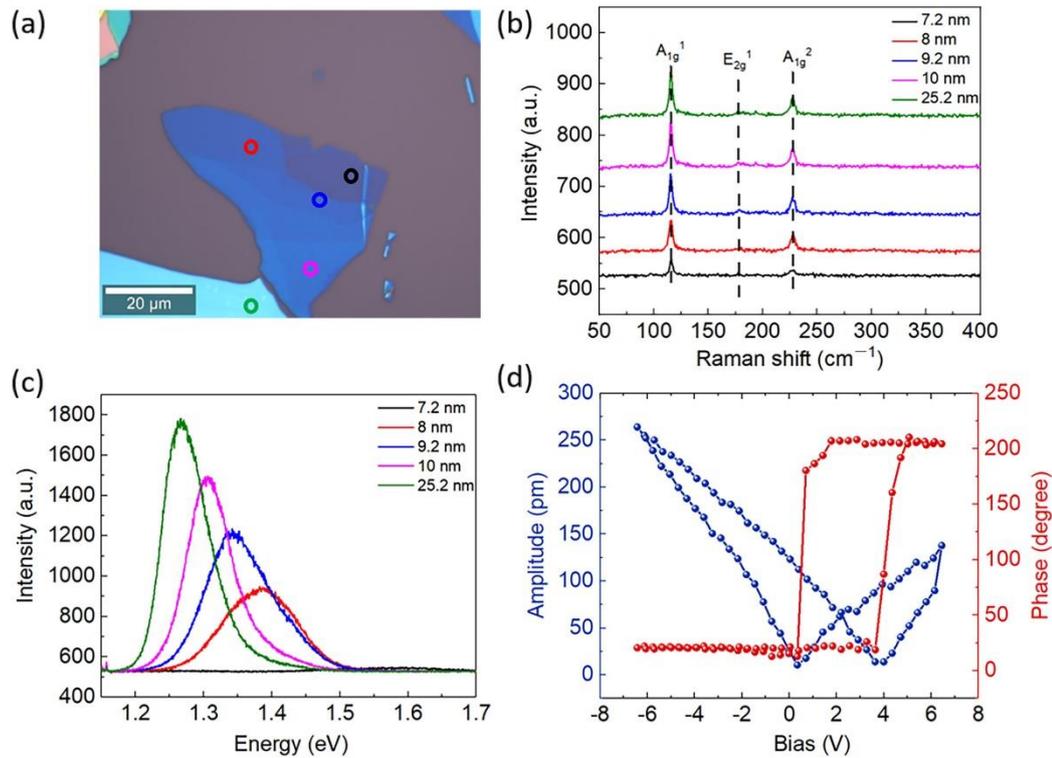

**Fig.1.** (a) Optical image of β-InSe nanosheet with different thickness. (b) Raman spectrum of β-InSe nanosheets with different thickness. (c) PL spectrum of β-InSe nanosheets with different thickness. (d) PFM amplitude and phase loop curve of 8-nm-thick β-InSe nanosheet.

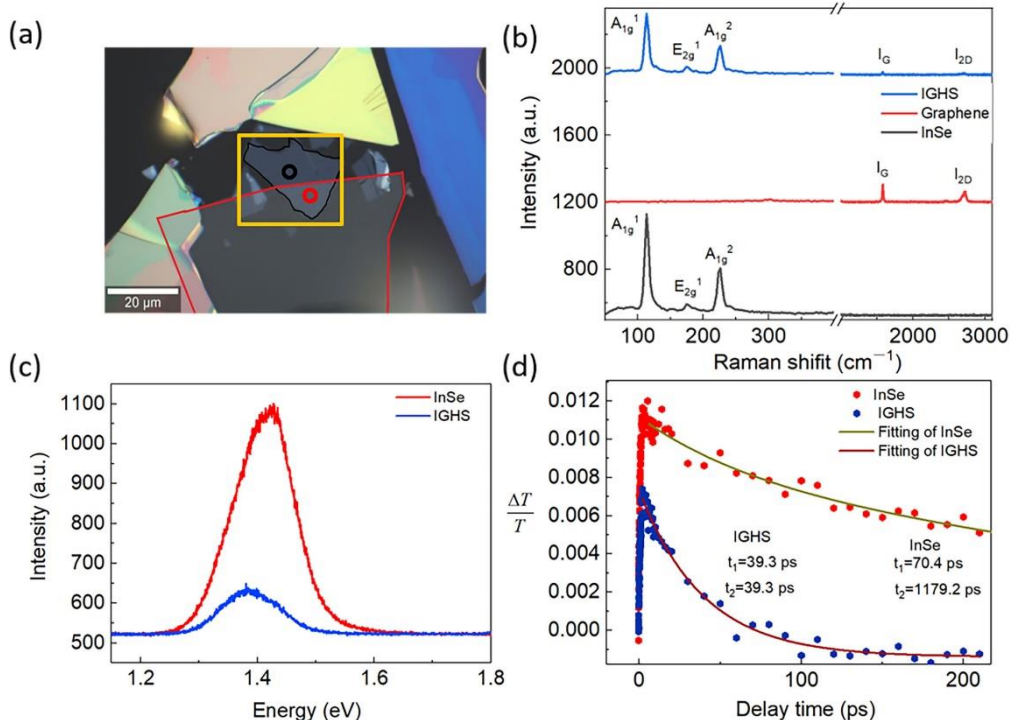

Fig.2. (a) The OM image (The yellow line is top InSe and red line is bottom graphene) of the fabricated IGHS sample. (b) Raman, (c) PL spectrum and (d) TA dynamics of InSe (black circle in (a)) and IGHS (red circle in (a)) respectively.

Fig. 3(a) shows the OM image of the fabricated in-plane GIGHS photodetector. Photolithography and metal depositing (5 nm Cr/25 nm Au) were used to fabricate the bottom electrodes, then the graphene flakes and β-InSe flake were transferred in sequence. For the decayed mobility problem of thin InSe flakes[19], the thickness of β-InSe chosen here is about 45 nm. To investigate the photoelectric performance of the fabricated in-plane GIGHS photodetector, we tested the photo response under different illumination power density with a continuous-wave laser of λ = 532 nm, as shown in Fig. 3(b). The tests were measured by using a Keithley 2450 sourcemeter without applying gate voltage. The relationship between photo current, responsivity, and incident laser power density with bias voltage of 1.5 V is shown in Fig. 3(c). The photocurrent of the photodetector is denoted as: $I_{ph} = I_{light} - I_{dark}$, the photo responsivity $R = I_{ph}/P$, where P is the incident laser power on the device area. As the illumination power density increasing, $I_{ph}$ is increasing while the responsivity decreases. As the laser power illuminating on the photodetector decreased to 9.17 pW, the photocurrent is 1.99 nA, thus $R = 2.17 \times 10^2$ A/W. We also estimate the maximum external quantum efficiency EQE = R(hv/e) ≈$5.06 \times 10^4$ %, demonstrating the high photo gain of the device. Furthermore, the detectivity is given by $D^* = RA^{1/2}(2eI_{dark})^{-1/2}$[11, 20], where e is the electron charge and A is the device channel area. Thus, detectivity ($D^*$) calculated is $3.78 \times 10^{13}$ Jones, and the corresponding noise equivalent power (NEP) value is $4.59 \times 10^{-17}$ W/Hz$^{1/2}$. The rising time of the photodetector (photocurrent arise from 10% to 90%) is 314 μs, and falling time (90% to 10%) is 308 μs, as shown in Fig. 3(d). It can be found that the photo responsivity and response time outperform that of individual InSe[16, 21-23], graphene[24], and many other 2D materials[26].

Here we discuss the mechanism of the high performance of our device. Firstly, the large photo gain and sublinear (fitting power exponents of 0.78) power-dependent photocurrent behavior indicates the mechanism of photogating effect, which the surface trapped carriers act as a local gate to boost the photocurrent[27]. Although the response time of other low-dimensional material photodetectors based on photogating mechanism is quite slow (>1s)[28, 29], the relatively short response time of our device should be attributed to the less trap states (the fitting power exponent of 0.78 is not far smaller than 1) and fast carrier relaxation time of the IGHS from our TA results. Secondly, due to the Schottky barrier at the interface, the photo-generated electrons and holes separate effectively, further resulting much lower dark current of 0.31 nA, and high detectivity of $3.78 \times 10^{13}$ Jones. Compared with the best performance of InSe (γ phase) photodetector with in-plane graphene electrodes[25], the β-InSe device shows one order higher detectivity than γ-InSe with lower bias voltage of 1.5 V . To further achieve a higher photo responsivity and faster response time, we also constructed an out-of-plane configuration device by transfer the β-InSe between two graphene flakes, the optimized device achieved responsivity of $2.12\times10^4$ A/W, detectivity of $1.73\times10^{14}$ Jones and response time of ~241 μs (Fig.S1), the superior performance is due to the effective interfacial transfer and shortened transmit distance for the photogenerated carriers. Consider all the parameters above, as Fig. 3(e) is shown, the balanced performance at zero-gate voltage is superior than the most reported low-dimensional Vis-NIR photodetectors, which is close to the idealized "targeted" performance region[28, 29].

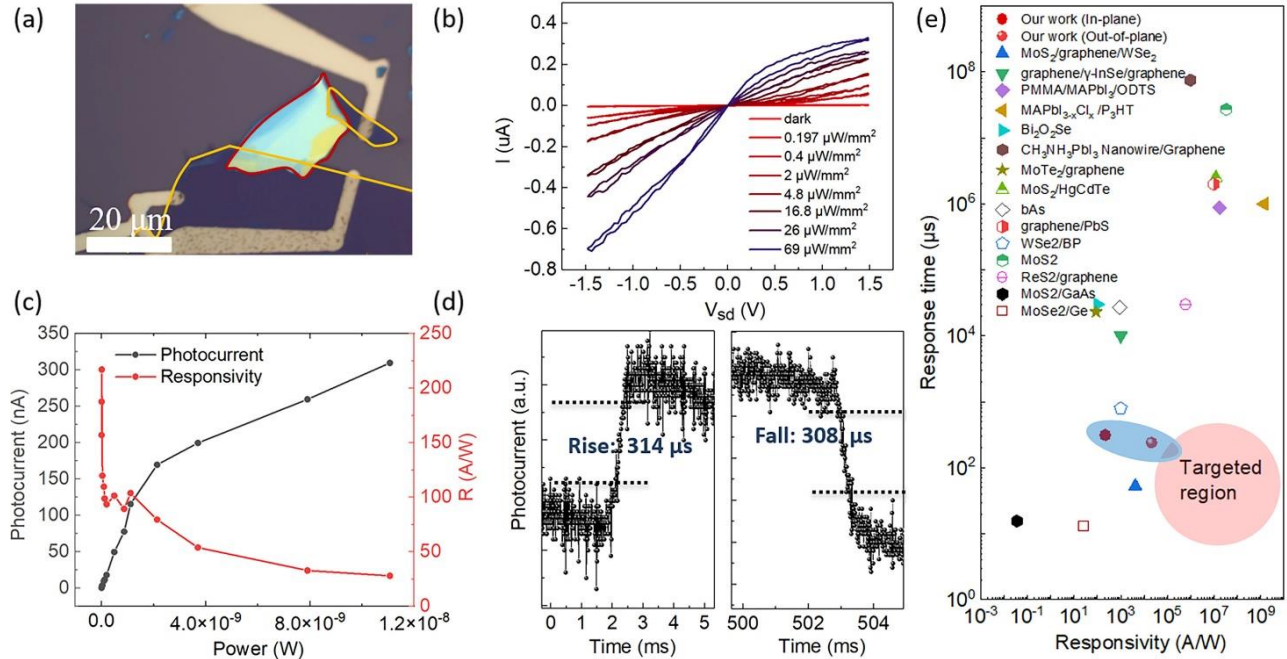

**Fig.3.** (a) The OM image of fabricated In-plane GIGHS photodetector device fabricated on a 285 nm-thick $SiO_2$-coated silicon substrate ((The yellow line is bottom graphene and red line is top β-InSe). The scale bar is 20 μm. (b) I−V curves of the IGHS photodetector with different power density at the wavelength of 532 nm under forward and backward bias voltage sweep. (c) Photocurrent and corresponding photo responsivity under bias voltage of 1.5 V. (d) The response time curve of the In-plane GIGHS photodetector device under the bias voltage of 1.5 V (λ = 532 nm). (e) Photo responsivity and response time of low-dimensional Vis-NIR photodetectors[25, 30])

Furthermore, to investigate the ferroelectricity effect on the photodetector performance, another in-plane configuration GIGHS device was fabricated, as shown in Fig. 4(a). To avoid sample breakdown problem under high voltage, the thickness of β-InSe chosen is about 12 nm here, due to the lower coercive field in thin flake for realizing ferroelectric switching[31]. The on and off states can be switched by applying the poling voltages of ±8 V, as shown in the I-V curves in Fig. 4(b). The photo responses affected by the poling process are presented in Fig. 4(c), which show the I–V curves of the photocurrent under 1 μW/mm² as a case. The calculated performance is compared in Table 1. The device after poling has a slightly lower responsivity than the unpoled, probably due to the induced traps caused by ferroelectric polarization field, which the photogating effect contribute less, the whole process is illustrated in Fig. 4(d), (e), (f). However, the detectivity under negative poling, has near six times higher compared with unpoled device under 2 V bias voltage, due to the lower dark current, resulting from an extra built-in potential driven by ferroelectric polarization field, as shown in Fig. 4(f).

In summary, photodetectors based on high-quality in-plane and out-of-plane GIGHS are fabricated and measured. High responsivity of $2.17\times10^2$ A/W, detectivity of $3.78\times10^{13}$ Jones, fast response time (314 μs) (In-plane configuration), and $2.12 \times 10^4$ A/W, detectivity of $1.73\times10^{14}$ Jones, fast response time (241 μs) (Out-of-plane configuration) are achieved under λ = 532 nm laser illumination at gate-free condition. The balanced performance is close to the idealized "targeted" performance region with both high responsivity and fast response time, due to the concurrent play of efficient interfacial carriers' transfer and photogating. Besides, the ferroelectricity effect on the photodetector performance was investigated. Compared to the unpoled device, the detectivity under negative poling has been improved by six times, due to the successful suppression of the dark current. Our work spotlights the potential application for fast sensitive opto-electronic applications by exploring high mobility ferroelectric semiconducting materials.

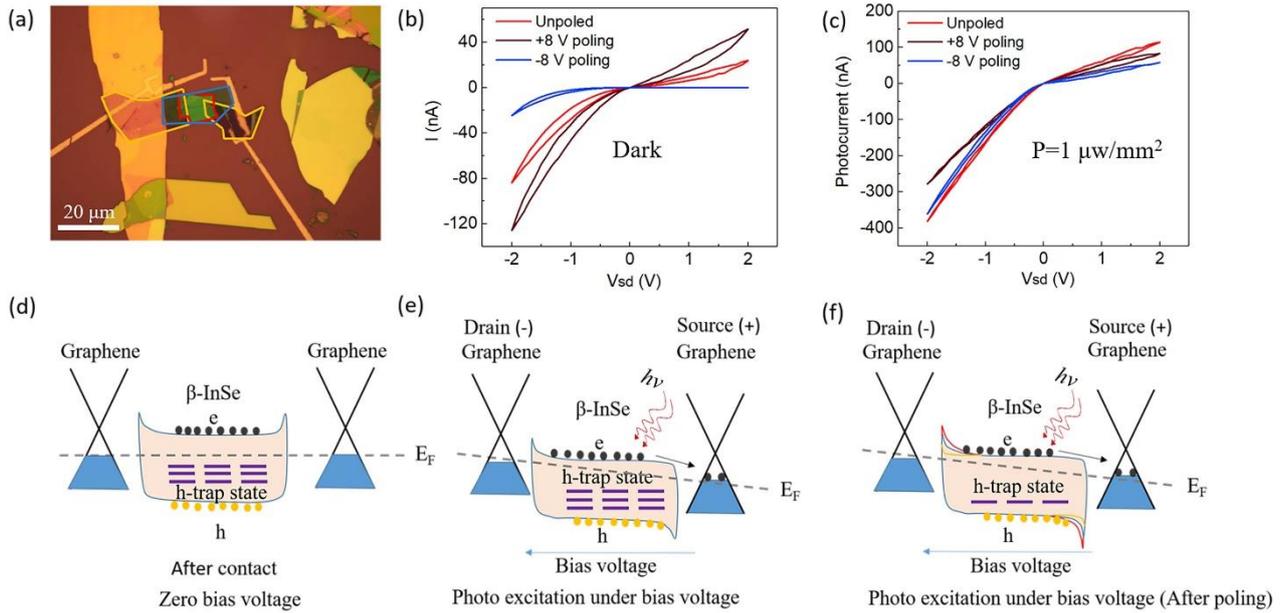

**Fig.4.** (a) The OM image of fabricated in-plane configuration GIGHS device ((The yellow line is bottom graphene, red line is middle β-InSe and blue line is top h-BN to completely isolate β-InSe from the environment). (b), (c) I–V curves of GIGHS device with the poling voltages of ±8 V under dark sate and light-on state respectively, relative to the reference case before poling. (d) Schematic energy band diagram of GIGHS without any bias voltage. (e) Schematic energy band diagram of GIGHS under photo excitation with bias voltage. (f) Schematic energy band diagram of GIGHS under photo excitation with bias voltage (After poling). Changes in the built-in potentials caused by the additional polarization field is highlighted by the red curves (negative poling) and yellow curves (positive poling).

**Table 1. Calculated photodetector performance under the poling voltages of ±8 V and unpoled for reference.**

| Poling voltage | Responsivity ($\times 10^4$ A/W) | | Detectivity ($\times 10^{13}$ Jones) | |
|---|---|---|---|---|
| | -2 V | 2 V | -2 V | 2 V |
| -8 Poled | 2.83 | 0.46 | 10.4 | 21 |
| +8 Poled | 2.18 | 0.65 | 3.44 | 1.6 |
| Unpoled | 2.99 | 0.9 | 5.78 | 3.24 |

See the supplementary material for the detailed sample preparation and optical characterization process, performance of out-of-plane configuration GIGHS photodetector device, and detailed calculation process for responsivity, detectivity, EQE and NEP calculation.

**Acknowledgments**

This work was supported by the funding from National Key R&D Program of China (2019YFA0308602), National Science Foundation of China (general program 12174336& major program 91950205) and the Zhejiang Provincial Natural Science Foundation of China (LR20A040002).

# Supporting Information

# Highly sensitive photodetector based on two-dimensional ferroelectric semiconducting β-InSe/graphene heterostructure

Jialin Li[1], Yuzhong Chen[2], Yujie Li[2], Haiming Zhu[2], Linjun Li[1,3]*

[1]State Key Laboratory of Modern Optical Instrumentation, College of Optical Science and Engineering, Zhejiang University, Hangzhou 310027, China
[2]Center for Chemistry of High-Performance & Novel Materials, Department of Chemistry, Zhejiang University, Hangzhou 310027, China
[3]Intelligent Optics & Photonics Research Center, Jiaxing Research Institute Zhejiang University, Jiaxing 314000, China

*Email: lilinjun@zju.edu.cn

# Experiment section

*Sample Preparation.* Our β-InSe crystals are obtained commercially (from SixCarbon Technology, Shenzhen) by using the Bridgman method. For fabrication of the nano-thick devices, traditional mechanic exfoliation method was used. Photolithography and metal depositing (5 nm Cr/25 nm Au) were used to fabricate the bottom electrodes of the IGHS photodetector, then the IGHS was transferred on the electrodes. The thickness of the sample is determined by atomic force microscopy under non-contact mode (Park NX-10). The MFP-3D infinity AFM was used to perform the PFM and hysteresis loops measurement. The spring constant of the stiff cantilever and AC voltage were set to 2.8 N/m and 0.5 V respectively.

*Optical Characterization.* Raman scattering and steady-state PL measurement on β-InSe single crystal is performed by a Raman spectrometer (Witec alpha300) with a 532 nm excitation laser source, a 100× objective is used in IGHS characterization. For femtosecond TR spectroscopy, the fundamental beam (λ=1030 nm, ~170 fs pulse duration) from Yb:KGW laser (Light Conversion Ltd Pharos) is separated to two paths, one is introduced into a noncollinear optical parametric amplifier to generate pump pulse at visible and near-IR wavelengths, while another is focused onto a YAG crystal after a delay-line to produce white light continuum (λ=500 ~ 950 nm) as probe light. A reflective 50× objective is used to focus both pump and probe beams onto the sample, which produces a focused spot size of ~2 μm. The resolution limit of our TA setup is about 34 fs. For photocurrent measurement, a commercial continuous-wave laser of

λ=532 nm is used as a light source, the photocurrent under biased voltage are measured by using a Keithley 2450 sourcemeter (Tektronix Inc.), and photo response time is recorded by a digital oscilloscope.

## Performance of out-of-plane configuration GIGHS photodetector device

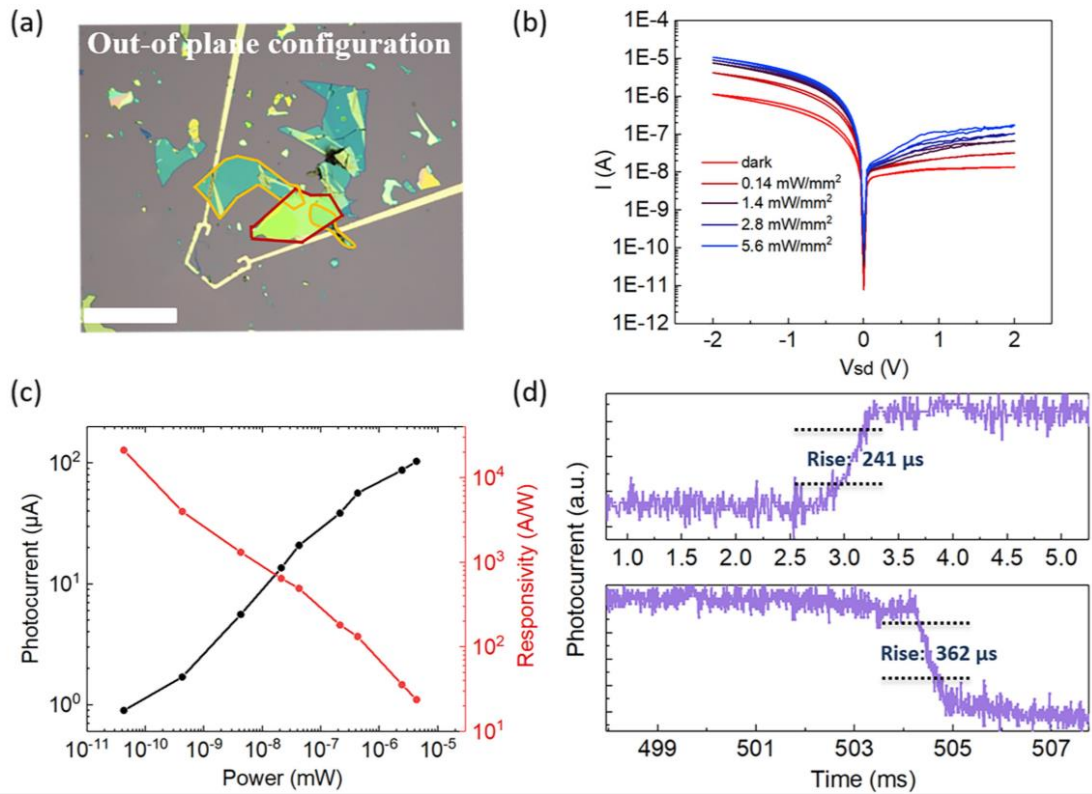

**Fig.S1.** (a) The OM image of fabricated out-plane GIGHS photodetector device (The yellow line is bottom graphene, red line is middle β-InSe and blue line is top graphene). The scale bar is 100 μm. (b) I−V curves of the out-plane GIGHS photodetector with different power density under forward and backward bias voltage sweep. (c) Photocurrent and corresponding photo responsivity of out-plane GIGHS photodetector under -2V bias voltage. (d) The response time curve of the out-of-plane GIGHS photodetector device under the bias voltage of -2 V (λ=532 nm).

## Supplementary Note

**Responsivity, detectivity, EQE and NEP calculation**

The responsivity of a photodetector can be expressed in units of amperes per watt of incident radiant power, defined as $R = I_{ph}/P$, where $I_{Ph}$ is the net photocurrent, P is the light power on the sample.

a. *For GIGHS with in-plane configuration ($V_{sd}$=1.5 V):*

The laser power on the sample is 9.17 pW, the net photocurrent is 1.99 nA, thus $R = 2.17\times10^2$ A/W.

The detectivity is given by $D^* = RA^{1/2} (2eI_{dark})^{-1/2}$, where e is the electron charge, $I_{dark}$ is the dark current and A is the device channel area. The unit of $D^*$ is ``Jones´´, 1 m $W^{-1}Hz^{1/2}$=100 cm $W^{-1}Hz^{1/2}$=100 Jones.

The $I_{dark}$ is 0.31 nA, device channel area is about 300 μm², thus $D^*=3.78\times10^{13}$ Jones.

The external quantum effifiency EQE = R [hv/e]=R [hc/λe], λ=532 nm, h is the planck constant, thus EQE=$5.06\times10^2$ (or $5.06\times10^4$ %).

The noise equivalent power (NEP) value is given by NEP=$S^{1/2}/D^*$ (*Adv. Funct. Mater.* 2020, 30, 1908427; *Laser Photonics Rev.* 2022, 2200338), thus NEP=$4.59\times10^{-17}$ W/$Hz^{1/2}$.

b. *For GIGHS with out-of-plane configuration ($V_{sd}$=-2 V):*

The laser power on the sample is 42.46 pW, the net photocurrent is 0.9 μA, thus $R = 2.12\times10^4$ A/W.

The $I_{dark}$ is 1.4 μA, device channel area is about 3000 μm², thus $D^*=1.73\times10^{14}$ Jones.

The external quantum effifiency EQE =R [hc/λe], λ=532 nm, h is the planck constant, thus EQE=$4.94\times10^4$ (or $5.06\times10^6$ %).
The noise equivalent power (NEP) value is given by NEP=$S^{1/2}/D^*$, thus NEP=$3.16\times10^{-17}$ W/$Hz^{1/2}$.

*c. Another GIGHS with in-plane configuration to investigate the ferroelectricity effect:*

After -8 V poling (Vsd=-2 V): The laser power on the sample is about 12.74 pW, the net photocurrent is 361 nA, thus R = $2.83 \times 10^4$ A/W. The $I_{dark}$ is 24.57 nA, device channel area is about 10 μm$^2$, thus D*=$1.01 \times 10^{14}$ Jones.

After -8 V poling (Vsd=+2 V): The laser power on the sample is about 12.74 pW, the net photocurrent is 58.38 nA, thus R = $4.58 \times 10^3$ A/W. The $I_{dark}$ is 0.149 nA, device channel area is about 10 μm$^2$, thus D*=$2.09 \times 10^{14}$ Jones.

After +8 V poling (Vsd=-2 V): The laser power on the sample is about 12.74 pW, the net photocurrent is 277.8 nA, thus R = $2.18 \times 10^4$ A/W. The $I_{dark}$ is 125.7 nA, device channel area is about 10 μm$^2$, thus D*=$3.44 \times 10^{13}$ Jones.

After +8 V poling (Vsd=+2 V): The laser power on the sample is about 12.74 pW, the net photocurrent is 83.4 nA, thus R = $6.5 \times 10^3$ A/W. The $I_{dark}$ is 51.67 nA, device channel area is about 10 μm$^2$, thus D*=$1.6 \times 10^{13}$ Jones.

Unpoled (Vsd=-2 V): The laser power on the sample is about 12.74 pW, the net photocurrent is 381.1 nA, thus R = $2.99 \times 10^4$ A/W. The $I_{dark}$ is 83.65 nA, device channel area is about 10 μm$^2$, thus D*=$5.78 \times 10^{13}$ Jones.

Unpoled (Vsd=+2 V): The laser power on the sample is about 12.74 pW, the net photocurrent is 114.04 nA, thus R =$8.95 \times 10^3$ A/W. The $I_{dark}$ is 23.85 nA, device channel area is about 10 μm$^2$, thus D*=$3.24 \times 10^{13}$ Jones.